\documentclass[preprint,aps,prb,showpacs,showkeys]{revtex4}
\usepackage{graphicx}
\usepackage{amssymb}
\usepackage{bm}
\begin{document}
\setcounter{page}{1}
\title[]{Oscillatory tunneling magnetoresistance in magnetic tunnel junctions
with inserted nonmagnetic layer }
\author{Changsik \surname{Choi}}
\author{Byung Chan \surname{Lee}}
\email{chan@inha.ac.kr}
\affiliation{Department of Physics, Inha University, Incheon
402-751, Republic of Korea}

\date{\today}

\begin{abstract}

Oscillatory tunneling magnetoresistance (TMR) as a function of
spacer thickness is investigated theoretically for a magnetic tunnel
junction with a nonmagnetic layer inserted between the tunnel
barrier and the ferromagnetic layer. TMR is characterized in an
analytical form, that is expressed with the transmission and
reflection amplitudes of single interfaces at the Fermi level, and
by the extremal wave vectors. Electronic structures with multiple
bands are taken into account in the derivation characterizing the
TMR, and the proposed analytical expression can be directly applied
to real junctions. Based on our model, the features of TMR
dependence on spacer thickness are discussed, including selection
rules for the oscillation period. Numerical calculations are
performed using an envelope-function theory for several cases, and
we show that our model is in good agreement with the exact result.

\end{abstract}



\maketitle

\section{INTRODUCTION}

Since high tunneling magnetoresistance (TMR) was first observed at
room temperature,\cite{mood} magnetic tunnel junctions (MTJs) have
been a focus of interest. Extensive research has been carried out to
understand and improve the properties of MTJs. A huge increase in
TMR with lower junction resistance was achieved when AlO$_x$ tunnel
barriers were replaced by MgO, and this was followed by realizations
of memory devices based on MTJs with MgO barriers.\cite{yuasaR}
Tunneling current in the MTJ is spin polarized, which adds another
dimension to the tunneling effect, and scientific attention has thus
been drawn to the spin-dependent tunneling phenomenon. When a
nonmagnetic (NM) layer is inserted between a ferromagnetic (FM)
layer and the insulating (I) tunnel barrier of the MTJ, the spin
polarization of the tunneling current changes and the TMR is
directly affected. An early theoretical work predicted the
oscillatory TMR as a function of the NM thickness due to the quantum
well states inside the NM layer.\cite{vedy} In sputtered samples, it
has been shown that an NM layer between the tunnel barrier and FM
layer could be detrimental to TMR, and the TMR decreases as a
function of NM thickness.\cite{dust,dust2} These experimental
results have been explained theoretically with a free electron
model, and the decay of TMR was attributed to a loss of coherence in
the electron propagation.\cite{zhang} Different experimental results
have been obtained for a crystalline NM layer inserted between the
tunnel barrier and the FM layer. Yuasa {\it et al.}\cite{yuasa}
experimentally investigated the dependence of TMR on Cu thickness in
NiFe/AlO$_x$/Cu/Co junctions with samples grown by molecular beam
epitaxy. They found that the TMR decayed but oscillated as a
function of Cu thickness. The oscillation period was determined by
the nesting feature of the Cu Fermi surface. This oscillatory TMR
has been investigated theoretically based on a single-band
tight-binding model, a free-electron model, and full-band
calculations.\cite{itoh, itoh2, shok, zeng, yang,  niu, feng, chen,
autes, chen2, chen3, lee1}  Many features have been explained with
calculations using simple models, but direct comparisons with the
experimental data are difficult because realistic electronic
structures were not considered. The full-band calculations are very
useful for the description of real systems. However they are time
consuming, and sometimes it is not easy to understand the underlying
physics. Furthermore, full-band calculations are usually carried out
for an ideal situation, and significant discrepancies often occur
between theory and experiments.

We introduce an analytical expression that describes the dependence
of TMR on NM thickness for FM/I/NM/FM junctions based on full-band
structures. Our approach uses the generalization of a previously
described single-band case\cite{lee1} to a multiple-band case that
considers the real materials. The TMR is expressed with transmission
and reflection amplitudes of single interfaces at the Fermi level,
and extremal wave vectors. The full-band structures of the materials
are taken into account in our proposed model, and the calculation of
several transmission and reflection amplitudes with real band
structures can make a direct comparison with experimental results
possible. Based on our model, selection rules for the oscillation
period are discussed, and we suggest that very few oscillation
periods will be observed in experiments even when there are many
extremal spanning vectors of the NM Fermi surface. This situation is
very different than the interlayer exchange coupling in magnetic
multilayers. Our model explicitly shows that TMR dependence on NM
thickness is affected by the thickness of the tunnel barrier, and
predicts that the TMR will go to zero as the NM thickness increases.
To check the validity of our model, we carried out numerical
calculations using an envelope-function theory for several cases
including NM material with a Fermi surface similar to the Cu(001)
case.
Although our model is calculated using the parameters at the Fermi
level and the extremal wave vector, it is in good agreement with the
exact numerical results.

\section{Theoretical Model}

Figure 1(a) shows a schematic diagram of an MTJ with an NM layer
inserted between the tunnel barrier and the right magnetic layer
[FM(R)]. The growth direction is taken as the $z$ axis, and $d$ is
the thickness of the NM layer. We used a frozen potential
approximation, and the eigenstate of the MTJ is expressed with
linear combinations of the bulk states for each layer. We assumed
that the wave vector component parallel to the interface (${\bf{k}
}_{\|}$) is conserved throughout the MTJ. A two-channel model was
adopted and spin-flip scattering was ignored. The normalized bulk
solution of the material for the left magnetic layer [FM(L)] is
denoted as $\left| \varepsilon, {\bf k}_{\|},k_{z,n \sigma}^{L+(-)}
\right\rangle$ for a given energy $\varepsilon$ and ${\bf{k}
}_{\|}$, where $k_z$ is the $z$ component of the wave vector, $n$ is
the band index, $\sigma$ is the spin index, and the $+$ $(-)$ sign
is for the state traveling to the right (left). Similarly, the bulk
solutions of the FM(R) and NM materials are expressed as $\left|
\varepsilon,  {\bf k}_{\|},k_{z,n \sigma}^{R+(-)} \right\rangle$ and
$\left| \varepsilon,  {\bf k}_{\|},k_{z,n }^{N+(-)} \right\rangle$,
respectively. Multiple bands are taken into account, and $2N_{L(R)
\sigma}$ is the number of bulk states in the left (right) FM layer
for a given $\varepsilon$, ${\bf{k} }_{\|}$, and spin $\sigma$. The
number of bulk states in the NM layer for a given $\varepsilon$ and
${\bf{k} }_{\|}$ is denoted as $2N_{N}$. The eigenstate of the MTJ
is written as
\begin{equation}
 \left| \psi_{\sigma}({\varepsilon,\bf
k}_{\|}) \right\rangle = \left\{
\begin{array}{ll}
\displaystyle \sum_{n=1}^{N_{L \sigma}} A_{n \sigma}^{+}
\frac{\left| \varepsilon,  {\bf k}_{\|},k_{z,n \sigma}^{L+}
\right\rangle } {\sqrt{\left| v_{z,n \sigma}^{L+} \right| }} +
\sum_{n=1}^{N_{L \sigma}} A_{n \sigma}^{-} \frac{\left| \varepsilon,
 {\bf k}_{\|}, k_{z,n \sigma}^{L-} \right\rangle }
{\sqrt{\left| v_{z,n \sigma}^{L-} \right| }} , & z <0,  \\
{\displaystyle \sum_{n=1}^{N_{N}}} C_{n \sigma}^{+}
                      \frac{\displaystyle \left| \varepsilon,  {\bf k}_{\|}, k_{z,n }^{N+} \right\rangle }
                      {\displaystyle \sqrt{\left| v_{z,n }^{N+} \right| }}
                      + {\displaystyle \sum_{n=1}^{N_{N}}} C_{n \sigma}^{-}
                      \frac{\displaystyle \left| \varepsilon,  {\bf k}_{\|}, k_{z,n }^{N-} \right\rangle }
                      {\displaystyle \sqrt{\left| v_{z,n }^{N-} \right| }}, & b < z < b+d,    \\
 \displaystyle     \sum_{n=1}^{N_{R \sigma}} B_{n \sigma}^{+}
                      \frac{\left| \varepsilon,  {\bf k}_{\|}, k_{z,n \sigma}^{R+} \right\rangle }
                      {\sqrt{\left| v_{z,n \sigma}^{R+} \right| }}
                      + \sum_{n=1}^{N_{R \sigma}} B_{n \sigma}^{-}
                      \frac{\left| \varepsilon,  {\bf k}_{\|}, k_{z,n \sigma}^{R-} \right\rangle }
                      {\sqrt{\left| v_{z,n \sigma}^{R-} \right| }}, & z > b+d,
                       \end{array}
                       \right.
\label{eq:MTJ}
\end{equation}
where $v_z$ is the $z$-component of the group velocity [$v_z =
(1/\hbar)(\partial \varepsilon /\partial k_z)$] for the
 corresponding bulk eigenstate, and $A^{\pm}_{n \sigma}$, $C^{\pm}_{n \sigma}$, and $B^{\pm}_{n \sigma} $
are coefficients to be determined from the boundary conditions. Note
that the bases in Eq.~(\ref{eq:MTJ}) are adjusted so that the
current is normalized. The eigenstate inside the tunnel barrier is
not shown here.
 We define vectors ${\bf
A}^{\pm}_{\sigma}$ and ${\bf B}^{\pm}_{\sigma}$ as ${\bf
A}^{\pm}_{\sigma} \equiv (A^{\pm}_{1 \sigma}, A^{\pm}_{2 \sigma},
\cdots , A^{\pm}_{N_{L \sigma} \sigma})^T$ and ${\bf
B}^{\pm}_{\sigma} \equiv (B^{\pm}_{1 \sigma}, B^{\pm}_{2 \sigma},
\cdots , B^{\pm}_{N_{R \sigma} \sigma})^T$. Then, ${\bf
A}^{-}_\sigma $ and ${\bf B}^{+}_\sigma$ are related to ${\bf
A}^{+}_\sigma $ and ${\bf B}^{-}_\sigma$ by the $S$-matrix,\cite{bruu}
\begin{equation}
\left( \begin{array}{c} {\bf A}^{-}_\sigma  \\ {\bf B}^{+}_\sigma
\end{array}     \right)
=\left( \begin{array}{cc} {\bf r}_\sigma & {\bf t}'_\sigma  \\
{\bf t}_\sigma  &  {\bf r}'_\sigma
\end{array}  \right)
\left( \begin{array}{c} {\bf A}^{+}_\sigma  \\ {\bf B}^{-}_\sigma
\end{array}     \right).
\label{eq:rt}
\end{equation}
Matrix element $t_{\sigma, nn'}$ ($r_{\sigma, nn'}$) is a kind of
transmission (reflection) amplitude for an incoming wave from the
left $\left| \varepsilon, \, {\bf
 k}_{\|},\, k_{z,n' \sigma}^{L+} \right\rangle$ to be transmitted (reflected) to
$\left| \varepsilon, \, {\bf k}_{\|},\, k_{z,n \sigma}^{R+}
\right\rangle $ ($\left| \varepsilon, \, {\bf k}_{\|},\, k_{z,n
\sigma}^{L-} \right\rangle $). The transmission (reflection)
amplitude for the opposite direction is given by $t'_{\sigma, nn'}$
($r'_{\sigma, nn'}$). We calculated the conductance $G$ for low bias
and zero temperature from the Landauer-B\"{u}ttiker formalism as
follows
\begin{equation}
G = \frac{e^2}{h}\sum_{{\bf{k}}_{\|},\sigma} {\mathrm {Tr}} \left[
{\bf t}_\sigma^{\dag} (\varepsilon_F,{\bf{k}}_{\|})  {\bf t}_\sigma
(\varepsilon_F,{\bf{k}}_{\|}) \right] , \label{eq:lan}
\end{equation}
where $\varepsilon_F$ is the Fermi energy.  As shown in Figs.~1(b) and
(c), we considered the FM(L)/I/NM and NM/FM(R) interfaces separately, and
expressed ${\bf t}_\sigma$ of the MTJ with the transmission and
reflection amplitudes of each separated interface. The eigenstate of
the FM(L)/I/NM system shown in Fig.~1(b) is expressed as
\begin{equation}
 \left| \psi^{L}_{\sigma}({\varepsilon,\bf
k}_{\|}) \right\rangle = \left\{
\begin{array}{ll} {\displaystyle
\sum_{n=1}^{N_{L \sigma}} A_{n \sigma}^{L+}} \frac{\displaystyle
\left| \varepsilon, {\bf k}_{\|}, k_{z,n \sigma}^{L+} \right\rangle
} {\displaystyle \sqrt{\left| v_{z,n \sigma}^{L+} \right| }} +
{\displaystyle \sum_{n=1}^{N_{L \sigma}}} A_{n \sigma}^{L-}
\frac{\displaystyle \left| \varepsilon, {\bf k}_{\|}, k_{z,n
\sigma}^{L-} \right\rangle }
{\displaystyle \sqrt{\left| v_{z,n \sigma}^{L-} \right| }} , & z <0,  \\
                        {\displaystyle \sum_{n=1}^{N_{N}}} C_{n \sigma}^{L+}
                      \frac{\displaystyle \left| \varepsilon,  {\bf k}_{\|}, k_{z,n }^{N+} \right\rangle }
                      {\displaystyle \sqrt{\left| v_{z,n }^{N+} \right| }}
                      + {\displaystyle \sum_{n=1}^{N_{N}}} C_{n \sigma}^{L-}
                      \frac{\displaystyle \left| \varepsilon,  {\bf k}_{\|}, k_{z,n }^{N-} \right\rangle }
                      {\displaystyle \sqrt{\left| v_{z,n }^{N-} \right| }}, & z >
                      b,
                       \end{array}
                       \right.
\end{equation}
where $A^{L\pm}_{n \sigma}$ and $C^{L\pm}_{n \sigma}$ are
coefficients.  We define ${\bf A}^{L\pm}_{\sigma} \equiv
(A^{L\pm}_{1 \sigma}, A^{L\pm}_{2 \sigma}, \cdots ,
A^{L\pm}_{N_{L_\sigma} \sigma})^T$ and ${\bf C}^{L\pm}_{\sigma}
\equiv (C^{L\pm}_{1 \sigma}, C^{L\pm}_{2 \sigma}, \cdots ,
C^{L\pm}_{N_{N} \sigma})^T$, and the relation between the
coefficients is expressed as
\begin{equation}
\left( \begin{array}{c} {\bf A}^{L-}_\sigma  \\ {\bf C}^{L+}_\sigma
\end{array}     \right)
=\left( \begin{array}{cc} {\bf r}_\sigma^L & {{\bf t}'}^{L}_\sigma  \\
{\bf t}_\sigma^L  &  {{\bf r}'}^L_\sigma
\end{array}  \right)
\left( \begin{array}{c} {\bf A}^{L+}_\sigma  \\ {\bf C}^{L-}_\sigma
\end{array}     \right).
\label{eq:rtL}
\end{equation}
Similarly, for the NM/FM interface shown in Fig.~1(c), the eigenstate is
given by
\begin{equation}
 \left| \psi^{R}_{\sigma}({\varepsilon,\bf
k}_{\|}) \right\rangle = \left\{
\begin{array}{ll}
\displaystyle \sum_{n=1}^{N_{N}} C_{n \sigma}^{R+}
                      \frac{\left| \varepsilon,  {\bf k}_{\|}, k_{z,n }^{N+} \right\rangle }
                      {\sqrt{\left| v_{z,n }^{N+} \right| }}
                      + \sum_{n=1}^{N_{N}} C_{n \sigma}^{R-}
                      \frac{\left| \varepsilon,  {\bf k}_{\|}, k_{z,n }^{N-} \right\rangle }
                      {\sqrt{\left| v_{z,n }^{N-} \right| }},
  & z <0,  \\
\displaystyle \sum_{n=1}^{N_{R \sigma}} B_{n \sigma}^{R+}
                      \frac{\left| \varepsilon,  {\bf k}_{\|}, k_{z,n \sigma}^{R+} \right\rangle }
                      {\sqrt{\left| v_{z,n \sigma}^{R+} \right| }}
                      + \sum_{n=1}^{N_{R \sigma}} B_{n \sigma}^{R-}
                      \frac{\left| \varepsilon,  {\bf k}_{\|}, k_{z,n \sigma}^{R-} \right\rangle }
                      {\sqrt{\left| v_{z,n \sigma}^{R-} \right| }},                         & z >
                      0.
                       \end{array}
                       \right.
\end{equation}
The vectors ${\bf C}^{R\pm}_{\sigma} = (C^{R\pm}_{1 \sigma},
C^{R\pm}_{2 \sigma}, \cdots , C^{R\pm}_{N_{N} \sigma})^T$ and ${\bf
B}^{R\pm}_{\sigma} = (B^{R\pm}_{1 \sigma}, B^{R\pm}_{2 \sigma},
\cdots , B^{R\pm}_{N_{R_\sigma} \sigma})^T$ are related as follows
\begin{equation}
\left( \begin{array}{c} {\bf C}^{R-}_\sigma  \\ {\bf B}^{R+}_\sigma
\end{array}     \right)
=\left( \begin{array}{cc} {\bf r}_\sigma^R & {{\bf t}'}^{R}_\sigma  \\
{\bf t}_\sigma^R  &  {{\bf r}'}^R_\sigma
\end{array}  \right)
\left( \begin{array}{c} {\bf C}^{R+}_\sigma  \\ {\bf B}^{R-}_\sigma
\end{array}     \right).
\label{eq:rtR}
\end{equation}
${\mathbf r}$ and ${\mathbf t}$ of the MTJ in Eq.~(\ref{eq:rt}) can
be expressed with ${\mathbf r}^L$, ${\mathbf t}^L$, ${{\mathbf
r}'}^L$, and ${{\mathbf t}'}^L$ in Eq.~(\ref{eq:rtL}) and ${\mathbf
r}^R$, ${\mathbf t}^R$, ${{\mathbf r}'}^R$, and ${{\mathbf t}'}^R$
in Eq.~(\ref{eq:rtR}) by considering the multiple reflection inside
the NM. We introduce the mean free path $\lambda$ due to scattering
inside the NM layer. For simplicity, we assumed that $\lambda$ is
constant, although the dependence of $\lambda$ on other parameters
can be included in our calculation. Then, the phase-coherent part of
the reflection amplitude ${\mathbf r}^c$ is given by
\begin{equation}
{\bf r}^c  = {\bf r}_\sigma^L + {{\bf t}'}_\sigma^L {\mathbf
\rho}_\sigma^R  {\mathbf \tau}_\sigma^L  e^{- \frac{2d}{\lambda}}
 + \sum_{n=1}^{\infty} {{\bf t}'}_\sigma^L ({\mathbf
\rho}_\sigma^R {{\mathbf \rho}'}_\sigma^L )^n {\mathbf
\rho}_\sigma^R  {\mathbf \tau}_\sigma^L  e^{- \frac{2d}{\lambda}
(n+1)},
\end{equation}
where the matrix elements $ \rho_{\sigma, nn'}^R$, $\tau_{\sigma,
nn'}^L$, and ${\rho'}_{\sigma, nn'}^L$ are   $ \rho_{\sigma, nn'}^R
= e ^{-i k_{z,n}^{N-}d} \,\, r_{\sigma, nn'}^R$,  $\tau_{\sigma,
nn'}^L = e ^{i k_{z,n}^{N+}d} \,\, t_{\sigma, nn'}^L$, and
${\rho'}_{\sigma, nn'}^L = e ^{i k_{z,n}^{N+}d} \,\, {r'}_{\sigma,
nn'}^L$. The phase-coherent part of the transmission amplitude ${\bf
t}^c$ can be obtained in a similar way. Because of the scattering
inside the NM layer, we have ${\mathrm {Tr}} [{{\bf t}^c}^{\dag}
{{\bf t}^c}] + {\mathrm {Tr}} [{{\bf r}^c}^{\dag} {\bf r}^c] <1 $
and we need to include the diffusive part of the transport. We
assume that the transmission back to the FM(L) layer through the
tunnel barrier is much smaller than that through the NM/FM(R)
interface. Then, the $1 -{\mathrm {Tr}} [{{\bf t}^c}^{\dag} {{\bf
t}^c}] - {\mathrm {Tr}} [{{\bf r}^c}^{\dag} {\bf r}^c]  $ portion
contributes to the sequential transmission.\cite{yang} Adding the
coherent and sequential transmissions, we have ${\mathrm {Tr}} [{\bf
t}^{\dag} {\bf t}] = 1- {\mathrm {Tr}} [{{\bf r}^c}^{\dag} {\bf
r}^c] $. Finally, using the properties of the $S$ matrix and taking
the first-order term in $e^{- {2d}/{\lambda}}$, we obtain
\begin{equation}
{\rm Tr} [{\bf t}^{\dag} {\bf t}] \cong {\rm Tr} [{{\bf t}^L}^{\dag}
{\bf t}^L] + 2 {\rm Re \, Tr}[{{\bf t}_\sigma^L}^\dag {{\bf
r}'}_\sigma^L {\mathbf \rho}_\sigma^R {\mathbf \tau}_\sigma^L ] e^{-
\frac{2d}{\lambda}}.
\end{equation}
Even when $\lambda$ is very large, this is a reasonable
approximation because the magnitude of the matrix element
$r_{\sigma, nn'}^R$ is less than 1, and the higher-order terms are
more rapidly oscillating as functions of $d$ and consequently
contribute less to the conduction. The conductance is given by
\begin{equation}
G = G_0 +  \frac{2e^2}{h}e^{- \frac{2d}{\lambda}} {\rm Re}
\sum_{{\bf{k}}_{\|},\sigma}{\rm  Tr}[{{\bf t}_\sigma^L}^\dag {{\bf
r}'}_\sigma^L {\mathbf \rho}_\sigma^R {\mathbf \tau}_\sigma^L ],
\end{equation}
where $G_0$ is the conductance of the FM(L)/I/NM junctions and the
energy is set to the Fermi level ($\varepsilon =\varepsilon_F$). The
conductance depends on the magnetic configurations, and we denote
the conductance for parallel (anti-parallel) magnetization of two
magnetic layers as $G_{\rm P(AP)}$. The TMR is given by $\Delta G/
G_{\rm AP}$, where $\Delta G$ is $\Delta G= G_{\rm P}-G_{\rm AP}$.
Here, we will show the calculation of $\Delta G$, and $G_{\rm
P(AP)}-G_0$ can be obtained in the same way. We define $\Delta
T_{nn'}^L$ and $ \Delta r^R_{nn'}$ as $\Delta T_{nn'}^L = \left(
{\bf t}^L_\uparrow  {{\bf t}^L_\uparrow}^\dag {{\bf
r}'}_{\uparrow}^L - {\bf t}^L_\downarrow {{\bf t}^L_\downarrow}^\dag
{{\bf r}'}_{\downarrow}^L \right)_{nn'} = \left| \Delta T_{nn'}^L
\right| e^{i \phi_{nn'}^L}$ and $ \Delta r^R_{nn'}= \left(
{r}^R_{\uparrow, nn'} - {r}^R_{\downarrow, nn'} \right)/2 = \left|
\Delta r^R_{nn'} \right| e^{i \phi_{nn'}^R}$, where $\uparrow$
($\downarrow$) is the majority (minority) spin. $\Delta G$ is
expressed as
\begin{eqnarray}
\Delta G &=&  \frac{4 e^2}{h} e^{- \frac{2d}{\lambda}}{\rm
Re}\sum_{n,n'} \sum_{{\bf{k}}_{\|},\sigma}
 \left| \Delta
T_{nn'}^L \right| \left| \Delta r^R_{nn'} \right| e^{i(q_{nn'}d +
\phi_{nn'})},
\end{eqnarray}
where $q_{nn'}$ and $\phi_{nn'}$ are $q_{nn'} = k_n ^{N+} - k_{n'}
^{N-}$ and $\phi_{nn'} = \phi_{nn'}^L+\phi_{nn'}^R$.

The summation over ${\bf{k}}_{\|}$ can be performed in a manner similar to
the calculation of the interlayer exchange coupling in magnetic
multilayers.\cite{lee,lee2} $\left| \Delta T_{nn'}^L \right|$ and
$e^{iq_{nn'}^F d}$ are rapidly changing as functions of ${\bf
k}_{\|}$. We assume the exponential dependence of $\left| \Delta
T_{nn'}^L \right|$ such that $\left| \Delta T_{nn'}^L \right| \propto
e^{-b \, \chi_{nn'} ({\bf{k}}_{\|} )}$.
Suppose that $ (k_{\alpha x} , k_{\alpha y} )$ is an extremal point,
which means $\nabla_{{\bf k}_{\|}} \left[-b \, \chi_{nn'} +
i(q_{nn'} d + \phi_{nn'}) \right] = 0$ at ${\bf k}_{\|}=(k_{\alpha
x} , k_{\alpha y} ) $. Since the main contribution to the integral
comes from the vicinity of the extremal point, we expand $-b \,
\chi_{nn'} + i(q_{nn'} d + \phi_{nn'}) $ around the extremal point
as follows
\begin{eqnarray}
-b \, \chi_{nn'} +i(q_{nn'}d + \phi_{nn'}) &\approx& -b \,
\chi_\alpha
+i(q_{\alpha} d + \phi_{\alpha}) \nonumber \\
& & - \left( \frac{b}{\kappa_{\alpha x}^b} - i \, \frac{d +
d_{\alpha x} }{\kappa_{\alpha x}^d} \right) (k_x - k_{\alpha x} )^2
\\
& & - \left( \frac{b}{\kappa_{\alpha y}^b} - i \, \frac{d +
d_{\alpha y} }{\kappa_{\alpha y}^d} \right)  (k_y - k_{\alpha y} )^2
, \nonumber  \label{e:expand}
\end{eqnarray}
where new parameters $\frac{1}{\kappa_{\alpha x}^b} = \frac{1}{2}
\frac{\partial^2 \chi_{nn'}}{\partial k_x^2} $,
$\frac{1}{\kappa_{\alpha y}^b} = \frac{1}{2} \frac{\partial^2
\chi_{nn'}}{\partial k_y^2} $, $\frac{1}{\kappa_{\alpha x}^d} =
\frac{1}{2} \frac{\partial^2 q_{nn'}}{\partial k_x^2} $,
$\frac{1}{\kappa_{\alpha y}^d} = \frac{1}{2} \frac{\partial^2
q_{nn'}}{\partial k_y^2} $, $d_{\alpha x}=\frac{1}{2} \kappa_{\alpha
x}^d \frac{\partial^2 \phi_{nn'}}{\partial k_x^2}$ and $d_{\alpha
y}=\frac{1}{2} \kappa_{\alpha y}^d \frac{\partial^2
\phi_{nn'}}{\partial k_y^2}$ are evaluated at $\varepsilon =
\varepsilon_F$ and ${\bf k}_{\|}=(k_{\alpha x} , k_{\alpha y} ) $.
Then, the summation over ${\bf k}_{\|}$ is carried out analytically
and $\Delta G$ is given by
\begin{equation}
\Delta G =  \frac{e^2}{h \pi} e^{- \frac{2d}{\lambda}}{\rm Re}
\sum_{\alpha}  \frac{n_\alpha
 \left| \Delta
T_{\alpha}^L \right| \left| \Delta r^R_{\alpha} \right| e^{
i(q_{\alpha} d + \phi_{\alpha} )}}{\sqrt{ \frac{b}{\kappa_{\alpha
x}^b} - i \, \frac{d + d_{\alpha x} }{\kappa_{\alpha x}^d} } \sqrt{
\frac{b}{\kappa_{\alpha y}^b} - i \, \frac{d + d_{\alpha y}
}{\kappa_{\alpha y}^d} }}
, \label{e:delG}
\end{equation}
where
$n_\alpha$ is the number of the extremal points of the same kind.
The phase of the square root is taken from $-\pi/2$ to $\pi/2$. The
parameters in Eq.~(\ref{e:delG}) are evaluated at the Fermi level
and the extremal point.
Suppose we have $\nabla_{{\bf k}_{\|}} \, \chi =0$ at ${\bf
k}_{\|}^b=(k_{ x}^b , k_{ y}^b )$ and $\nabla_{{\bf k}_{\|}} (qd +
\phi )=0$ at ${\bf k}_{\|}^d=(k_{ x}^d , k_{ y}^d )$. In general,
${\bf k}_{\|}^b$ is different from ${\bf k}_{\|}^d$ and the
corresponding extremal point $(k_{\alpha x},k_{\alpha y})$ is a
complex number. This makes other parameters such as $q_\alpha$
complex numbers, and the situation is rather complicated. However,
when ${\bf k}_{\|}^b$ and ${\bf k}_{\|}^d$ are far apart, the
contribution is negligible because of small $\left| \Delta
T_{\alpha}^L \right|$. The most important case is when ${\bf
k}_{\|}^b$ and ${\bf k}_{\|}^d$ coincide. This is expected to happen
often at ${\bf{k}}_{\|}=0$ due to symmetry. In this case,
$k_{\alpha x}$ and $k_{\alpha y}$ are real and $\Delta G$ becomes
\begin{equation}
\Delta G =  \frac{e^2 n_\alpha }{2 h \pi } {\rm Re} \frac{ \left(
 |t^L_\uparrow|^2 - |t^L_\downarrow|^2
\right) \left|  r^R_\uparrow - r^R_\downarrow \right|e^{-
\frac{2d}{\lambda}} e^{ i(q_{\alpha} d + \phi_{\alpha} )}}{\sqrt{
\frac{b}{\kappa_{\alpha x}^b} - i \, \frac{d + d_{\alpha x}
}{\kappa_{\alpha x}^d} } \sqrt{ \frac{b}{\kappa_{\alpha y}^b} - i \,
\frac{d + d_{\alpha y} }{\kappa_{\alpha y}^d} }}
, \label{e:delGsingle}
\end{equation}
where we used $r'^L_\uparrow \cong r'^L_\downarrow$ with
$|r'^L_\uparrow| \cong 1$, which is expected for a typical thickness
of the tunnel barrier. In this case, $q_\alpha$ is exactly the
extremal spanning vector of the NM Fermi surface.

The extremal spanning vector $q_\alpha$ of the NM Fermi surface
gives rise to the period of the TMR oscillation. In principle,
multiple periods are possible depending on the shape of the NM Fermi
surface. However, the extremal spanning vectors of the NM Fermi
surface are the necessary condition for TMR oscillation, and the
period would be observed only when the corresponding $\left| \Delta
T_{\alpha}^L \right|$ is large enough. Except for the symmetry point
in ${\bf{k}}_{\|}$ space, the extremal points of the NM Fermi
surface hardly coincide with the maxima point of $\left| \Delta
T_{\alpha}^L \right|$ of the tunnel barrier. Thus, only few periods
would be observable among the possible periods from the extremal
spanning vectors. Compared with the oscillation of the interlayer
exchange coupling for a corresponding spacer, fewer periods would be
observed in TMR oscillation. Moreover, as $b$ increases, $\left|
\Delta T_{\alpha}^L \right|$ decreases rapidly as a function of
${\bf{k}}_{\|}$ away from the maxima. Thus, the multiple periods
would be more difficult to observe for a thicker tunnel barrier.
Since $\left| \Delta T_{\alpha}^L \right|$ is material dependent,
the period of TMR oscillation could change for a different tunneling
barrier. In Eq.~(\ref{e:delGsingle}), the extra phase factor arises
from the square root in the denominator. This extra phase factor is
$d$ dependent, and the measured oscillation period will be slightly
different from $2 \pi/q_\alpha$. For instance, the period of TMR
oscillation for NiFe/AlO$_x$/Cu(001)/Co is expected to be slightly
smaller than $2 \pi/q_\alpha$ , which is the period of the
interlayer exchange coupling for fcc Co/Cu(001)/Co multilayers.

We considered the observable periods with the assumption that the
tunneling probability is maximum at ${\bf{k}}_{\|}=0$. Then, only
the oscillation for the extremal spanning vector at
${\bf{k}}_{\|}=0$ would be observed. In NiFe/AlO$_x$/Cu(001)/Co
junctions, there are two possible periods from the Fermi surface of
Cu along the (001) direction. The long period is from the extremal
spanning vector at ${\bf{k}}_{\|}=0$ , and the extremal vector for
the short period is far away from ${\bf{k}}_{\|}=0$. Thus, the short
period is intrinsically invisible, and only the long period could be
observed. There are four possible oscillation periods in
NiFe/AlO$_x$/Cu(110)/Co junctions from the Cu Fermi surface along
the (110) direction.\cite{lee} Among them, the only observable
period is the extremal spanning vector corresponding to
${\bf{k}}_{\|}=0$. This period is very short and would be easily
wiped out by the interface roughness. Thus, it would be difficult to
observe a TMR oscillation in this system. There is only one possible
period for Cu(111)\cite{lee,lee2} and this period has been observed
experimentally in the interlayer exchange coupling of Co/Cu(111)/Co
multilayers.\cite{stil} However, this period would not be observed
in NiFe/AlO$_x$/Cu(111)/Co junctions because the corresponding
extremal spanning vector is far from ${\bf{k}}_{\|}=0$. Therefore,
except for the long period of NiFe/AlO$_x$/Cu(001)/Co junctions, it
would be difficult to observe a TMR oscillation as a function of Cu
thickness. The situation for the Au and Ag spacers will be same as
in the case of Cu because the shapes of the Fermi surfaces are
similar. In the Fe/Cr/Fe multilayers, the long periods of interlayer
coupling oscillation have been observed as a function of the Cr
thickness for the (100), (110), and (211) orientations.\cite{stil}
However, these long periods would not be observed experimentally in
Fe/AlO$_x$/Cr/Fe or Fe/MgO/Cr/Fe tunnel junctions because the long
periods are from the $N$ point of the Cr Fermi
surface,\cite{tse,you} which is away from ${\bf{k}}_{\|}=0$ for any
orientation. In experiments, this long period has not been observed
for Fe/AlO$_x$/Cr(001)/Fe (Ref.~\onlinecite{naga}) and
Fe/MgO/Cr(001)/Fe (Ref.~\onlinecite{mats}) tunnel junctions.
Usually, the long period is clearly observed because it is not
eliminated by the interface roughness. Although we discussed the
case that the tunneling is dominated by perpendicularly incident
electrons, the analysis is similar when the tunneling probability is
maximum or high away from ${\bf{k}}_{\|}=0$. If the dependence of
the tunneling probability on ${\bf{k}}_{\|}$ changes with a
different tunnel barrier, the TMR dependence on the NM spacer will
be altered accordingly. Still, the crucial criterion is whether the
tunneling probability is significant or not at the point of the
extremal spanning vectors of the NM Fermi surface in the
${\bf{k}}_{\|}$ space. Except special points, the chances are that
the point of the maximum tunneling probability does not coincide
with the position of the extremal spanning vectors of the NM Fermi
surface in the ${\bf{k}}_{\|}$ space. Thus many oscillations
inferred from the NM Fermi surface would not be observed. When the
tunnel barrier is extremely thin, the situation can be much
different because the tunneling probability dependence on
${\bf{k}}_{\|}$ may change significantly. More oscillation periods
can be observed with thinner tunnel barriers. Even in this case, the
oscillation periods associated with relatively high tunneling
probability will be observed.

Without the scattering effect, $\Delta G$ and TMR decays as $1/d$
for a thick NM layer. However, for thin NM layers, the decay rate is
slower than $1/d$ and is affected by the tunnel barrier thickness
$b$, and also by $d_{\alpha x}$ and $d_{\alpha y}$ (${\bf{k}}_{\|}$
dependence of the reflection-amplitude phase factors). The amplitude
decays much more slowly than $1/d$ for coherent transport when $d$
is of the same order of magnitude as $b$. In experiments, the TMR
oscillation decays much faster than $1/d$, which seems to be due to
scattering. As the NM thickness increases, our model predicts that
$\Delta G$ and TMR go to zero even when the mean free path
($\lambda$) is very long. We will address this point in Sec.~IIIC.


\section{Numerical calculation with an envelope-function theory}

To test the validity of our model, we carried out numerical
calculations based on an envelope-function  theory for several
cases. We used the same material for FM(L) and FM(R), and ignored
scattering.  The continuity of the wave function and the
conservation of current at the interface were taken as the boundary
conditions.

\subsection{Effective-mass approximation}
 First, we considered the case that the dispersion
relation of the NM material is the same as that of the FM material
for majority spin. The dispersion relations of the FM material and
the insulator are given by $\varepsilon ({\rm \bold k})= \hbar^2
k^2/2m^*_\sigma + V_\sigma$ and $\varepsilon ({\rm \bold k})=
\hbar^2 k^2/2m_0 + V_I$, respectively, where $m^*_\sigma$ is the
spin-dependent effective mass of the FM material, $m_0$ is the bare
electron mass, and $V_I$ is the height of the tunnel barrier. We set
$V_\uparrow  = 0$ [$V_\downarrow  = \Delta$] for the majority
(minority) spin in the FM layer using the spin-splitting energy
$\Delta$. The effective mass of the FM material is $m_\uparrow =
m_0$ for the majority spin and $m^*_\downarrow = m_0 \varepsilon_F /
(\varepsilon_F - \Delta )$ for the minority spin. Schematics for the
dispersion relations of the FM, I, and NM materials are shown in
Fig.~2.  Under these conditions, all the traveling states in the NM
layer have corresponding traveling states in the FM layer, and total
reflection does not occur at the NM/FM interface for any ${\rm
\mathbf k}_{\|}$. The parameters used in the calculation are
$\varepsilon_F = 4$ eV, $V_I = 6$ eV, and $\Delta = 2.5$ eV. The TMR
is plotted as a function of NM thickness $d$ for the tunnel-barrier
thicknesses of $b=1$ nm and $b=2$ nm in Figs.~3(a) and 3(b),
respectively. The solid line represents the exact calculation, which
was obtained using Eq.~(\ref{eq:lan}). The dotted line is the result
of our analytical simple model described by
Eq.~(\ref{e:delGsingle}). Note that the extremal point for the NM
Fermi surface coincides with the maximum transmission point of the
tunnel barrier at ${\rm \bold k}_{\|}=0$. When the NM layer is thin,
there is some discrepancy between the exact result and our
analytical model, but the agreement improves as the NM thickness
increases. The overall trend of the NM thickness dependence is well
depicted by our analytical model. The TMR oscillates and goes to
zero as the NM thickness increases. It is also shown that the TMR
dependence on $d$ is affected by the thickness of the tunnel
barrier.
The peak points of TMR do not coincide for different tunnel barrier
thicknesses. The decay of TMR as a function of $d$ is faster for a
thinner tunneling barrier. It is essential to consider the ${\rm
\bold k}_{\|}$ dependence of the transmission coefficient in the
calculation. The effect of the tunnel barrier thickness on the NM
thickness dependence of TMR is well described by our proposed
analytical formula Eq.~(\ref{e:delGsingle}).


\subsection{Spacer with nonparabolic dispersion relation}

Second, to investigate the case of the multiple extremal spanning
vectors in the NM layer, we assumed the following effective
dispersion relation for the NM layer:
\begin{equation}
\varepsilon = \frac{1}{1-a^2} \left[ \left(\frac{\hbar^2 k_{\|}^2}{2
m_0} - a^2 \varepsilon_F \right)^2 + \left(\frac{\hbar^2 k_z^2}{2
m_0} \right)^2 \right]^{1/2}, \label{e:NM}
\end{equation}
where the constant $a$ is set to $a=0.68$. Except for the NM layer
and $b=1.5$ nm, the dispersion relations and parameters for the FM
and tunnel barrier are the same as in the previous case. The
effective mass and $k_z$ in the NM layer were determined using
Eq.~(\ref{e:NM}). In Fig.~\ref{NM}(a), the cross section of the
Fermi surface for bulk NM is plotted as a function of $k_{\|}$.
There are two kinds of extremal spanning vectors : one at $k_{\|}=0$
and the other at $k_{\|}=0.68 k_F$, where $k_F=\sqrt{2m_0
\varepsilon_F}/\hbar$ is the magnitude of the Fermi wave vector for
the FM with majority spin. The extremal spanning vector at
$k_{\|}=0$ ($k_{\|}=0.68 k_F$) is shorter (longer) and can give rise
to a long (short) period of the TMR oscillation, which is similar to
the Cu(001) case. The TMR as a function of $d$ is shown in
Fig.~\ref{NM}(b). The solid line is the exact calculation and the
dotted line is based on Eq.~(\ref{e:delGsingle}). The agreement is
fairly good, and only oscillation with a long period was observed.
For the analytical model calculation, we included only the extremal
spanning vector at $k_{\|}=0$ and ignored the contribution from the
extremal point at $k_{\|}=0.68 k_F$. This is because the tunneling
probability decreases rapidly away from $k_{\|}=0$ and the spin
asymmetry of the transmission coefficient ($|t^L_\uparrow|^2 -
|t^L_\downarrow|^2$) is very small at $k_{\|}=0.68 k_F$. The
observed behavior clearly shows that even though there are two
possible periods of the TMR oscillation from the Fermi surface of
the NM, only the period with significant spin asymmetry of the
transmission coefficient would survive. The case of multiple
extremal spanning vectors in the NM layer is also described well by
our analytical formula.

\subsection{Tunnel barrier with nonparabolic dispersion relation}

Third, we studied the case that the point for the maximum tunneling
probability does not coincide with the position of the extremal
spanning vector of the NM Fermi surface in ${\rm \bold k}_{\|}$
space.  We assumed that the dispersion relation of the tunnel
barrier is given by
\begin{equation}
\varepsilon =  \left[ \frac{\hbar^4 (k_{x}^2-a^2 k_F^2)^2}{4 m_0^2}
 +  \frac{\hbar^4 (k_{y}^2-a^2 k_F^2)^2}{4 m_0^2}+ V_{\|}^2 \right]^{1/2}
 + \frac{\hbar^2 k_z^2}{2 m_0} + V_B - V_{\|} ,
\label{e:B}
\end{equation}
where $V_B$ is the bottom energy of the tunnel barrier, $V_{\|}$ is
an energy parameter, and $k_F$ is the magnitude of the Fermi wave
vector for the NM material. With this tunnel barrier, the tunneling
probability is highest at ${\rm \bold k}_{\|} = (a k_F, a k_F) $.
For the calculation, we used $b=1.5$ nm, $V_B = 6$ eV, $V_{\|}=0.7$
eV, and $a=0.566$. The other materials are assumed to be the same as
in Sec.~III-A. The dispersion relation of the FM material is given
by $\varepsilon ({\rm \bold k})= \hbar^2 k^2/2m^*_\sigma + V_\sigma$
with $V_\uparrow  = 0$, $V_\downarrow  = \Delta$, $m_\uparrow =
m_0$, and $m^*_\downarrow = m_0 \varepsilon_F / (\varepsilon_F -
\Delta )$. The dispersion relation of the NM material is the same as
that of the FM material for majority spin. Note that the extremal
spanning vector of the NM Fermi surface is located at ${\rm \bold
k}_{\|}=0$. The parameters used in the calculation are
$\varepsilon_F = 4$ eV and $\Delta = 2.5$ eV. The TMR dependence on
the thickness of the NM layer is displayed in Fig.~5. The solid line
is the exact result and the dotted is the result of our analytical
formula. As expected, the TMR decays faster and it is almost
negligible when the NM layer is thicker than about 2 nm. The inset
is the transmission coefficient as a function of ${\rm \bold
k}_{\|}$ along the [110] direction for the FM/I/NM system. It has
the maximum value around $k_{110} = 0.8 k_F$, at which we have
$\nabla_{{\bf k}_{\|}} \left( \chi \right) =0$. On the other hand,
the spanning vector ($q$) of the NM Fermi surface has the longest at
${\rm \bold k}_{\|}=0$. The maximum point of the tunneling
probability does not coincide with the extremum of the NM Fermi
surface, and the extremal point is determined from $\nabla_{{\bf
k}_{\|}} \left[-b \, \chi + i(q d + \phi) \right] = 0$. The extremal
point is $d$-dependent and a complex number.
It is not easy to calculate the $d$-dependent extremal point exactly
and we used the following approximation. At the Fermi level, $k_z$
in Eq.~(\ref{e:B}) is an imaginary number ($k_z = i \kappa$). We
used a parabolic function of $\kappa$ which was expanded in a Taylor
series around ${\rm \bold k}_{\|} = (a k_F, a k_F) $. The
calculation became much simpler, and the extremal point and the
corresponding vector $q_{\alpha}$ in Eq.~(\ref{e:expand}) were
obtained immediately as functions of $d$. In the range of thin NM
layers with significant TMR, $q_{\alpha}$ is shorter than $2k_F$,
the extremal spanning vector of the NM Fermi surface. Thus, the
oscillation period in this region is longer than what is expected
from the NM Fermi surface. Whenever the extremal spanning vectors of
the NM Fermi surface are significantly away from the point of the
maximum transmission coefficient in the ${\rm \bold k}_{\|}$ space,
we expect rapid decay of TMR as a function of the NM thickness.

\subsection{Free-electron model}

Finally, we investigated the free-electron model case, where  all
the effective masses are simply the bare electron mass $m_0$. The NM
band is assumed to be the same as the majority-spin band of FM,
which is commonly adopted in the theoretical calculations. The
parameters used in the calculation are $\varepsilon_F = 4$ eV, $V_I
= 6$ eV, $\Delta = 3.5$ eV, and $b=1$ nm. A plot of the TMR
dependence on NM thickness is displayed by a dotted line in Fig.~6.
In this case, the TMR reaches a finite value when the NM thickness
becomes infinite. At first sight, this seems to contradict our
analytical formula Eq.~(\ref{e:delGsingle}) given in Sec.~II. The
reason for finite TMR at infinite $d$ can be explained as follows.
We obtained $k_{z} = \sqrt{2m_0 (\varepsilon_F - \Delta)/\hbar^2
-k_{\|}^2 } $ from the dispersion relation at the Fermi level.
When the magnetizations of the FM layers are antiparallel, electrons
with the majority spin in the FM(L) layer do not penetrate to the
FM(R) layer for $k_{\|} > \sqrt{2m_0 (\varepsilon_F - \Delta)}
/\hbar$ because $k_z$ becomes imaginary in the FM(R) layer. The
electrons with the majority spin for $k_{\|} > \sqrt{2m_0
(\varepsilon_F - \Delta)} /\hbar$ in the FM(L) layer are totally
reflected at the NM/FM(R) interface due to the potential step and do
not contribute to conduction. Thus, $G_{\rm AP}$ is underestimated
and gives rise to a finite $\Delta G$ and consequently a finite TMR,
even when the NM layer is infinite, as long as the transport is
coherent. This is more pronounced for a larger $\Delta$, lower
barrier height, and thinner tunneling barrier because a larger
portion of electrons with the Fermi energy will be completely
reflected at the NM/FM(R) interface. When deriving
Eq.~(\ref{e:delG}) in Sec.~II, we considered multiple reflection
inside the NM layer to calculate the transmission, and we assumed
that once the electrons tunnel through the tunnel barrier, most of
them flow to the FM(R) layer after multiple reflections. However, in
the AP magnetizations of FM layers of the free-electron model, the
electrons with majority spin for $k_{\|} > \sqrt{2m_0 (\varepsilon_F
- \Delta)} /\hbar$ in the FM(L) layer cannot penetrate into the
FM(R) layer after tunneling because of the potential step, resulting
in a finite TMR for the infinite NM layer. To clarify this point, we
calculated the TMR for the finite FM(R) layer; namely, the
FM(L)/I/NM/FM(R)/NM junctions. The FM(R) layer is 1 nm thick, and
the result is shown by the solid line in Fig.~6. The FM(R) layer
behaves like a potential barrier for electrons with the majority
spin and $k_{\|} > \sqrt{2m_0 (\varepsilon_F - \Delta)} /\hbar$ in
the FM(L) layer. However, the TMR goes to zero as the NM thickness
increases. When the FM(R) layer is not too thick, the tunneling
probability through this potential barrier is larger than that
through the tunnel barrier (I), and most of electrons with majority
spin and $k_{\|} > \sqrt{2m_0 (\varepsilon_F - \Delta)} /\hbar$ in
the FM(L) layer eventually flow to the NM(R) layer once they tunnel
through the insulating barrier. Even for the infinite FM(R) layer,
the TMR for the infinite NM layer becomes negligible as $\Delta$
increases, the thickness of the tunnel barrier increases, and the
barrier height increases. This is because only a small portion of
electrons with the Fermi energy will be completely reflected at the
NM/FM(R) interface. Also, if the minority spin band of the FM layer
is set to be the same as the NM band, the TMR goes to zero as the NM
thickness increases because there is no total reflection at the
NM/FM(R) interface. The finite TMR for the infinite NM layer is
possible when a significant portion of electrons with the Fermi
energy have less transmission probability into the FM(R) layer than
that through the tunnel barrier. Complete reflection of electrons at
the interface between two metals may occur for some given ${\rm
\bold k}_{\|}$ at the Fermi level due to the mismatch of the
electronic states. However, even slight scattering would lead to the
penetration of electrons into the FM(R) layer. In this sense, we do
not expect that a large portion of electrons with Fermi energy are
reflected completely, in reality, at the NM/FM interface. Thus, it
is unrealistic to expect a finite TMR as the NM thickness increases
in experiments.



\section{CONCLUSION}

We calculated the TMR of FM/I/NM/FM tunnel junctions. The TMR was
calculated as a function of NM thickness using the
Landaur-B\"{u}ttiker formula. Multiple band structures were included
and an analytical form describing the TMR was obtained. Conductance
was calculated from the summation of the transmission over ${\rm \bf
k}_{\|} $. The transmission was obtained by considering multiple
reflections between the I/NM and NM/FM interfaces. The summation
over ${\rm \bf k}_{\|} $ was carried out analytically. The
contribution was mainly from the extremal point in ${\rm \bf
k}_{\|}$-space that was determined from the combination of the NM
Fermi surface and the ${\rm \bf k}_{\|} $-dependence of the
transmission coefficient of the FM/I/NM junction. The TMR was
expressed with the transmission coefficient of the FM/I/NM junction,
reflection amplitudes at the NM/I and NM/FM interfaces at the Fermi
level, and the extremal wave vector. Many oscillation periods can be
inferred from the shape of the NM Fermi surface, but they can be
observed only when the corresponding spin asymmetry of the
transmission coefficient is significant. We suggest that only few
oscillation periods are likely to be observed in real experiments.
When the NM spacer is thin, our proposed model indicates that the
decay of the TMR was slower than the inverse of the space thickness
for coherent transport.

Numerical calculations were performed to investigate the accuracy of
the proposed formula. An envelope-function theory was adopted, and
our model was compared to the exact result. We showed that the
results of the proposed formula are in good agreement with the exact
calculations. The TMR dependence on the thickness of the NM spacer
was affected by the tunnel barrier thickness, which was well
described by our formula. The numerical calculation was extended to
the case with multiple extremal spanning vectors in the Fermi
surface of the NM spacer. Our proposed formula is in good agreement
with the exact result, and only the oscillation period with
significant spin asymmetry of the transmission coefficient was
observed as predicted using our formula. When the tunneling
probability associated with each extremal spanning vectors of the NM
Fermi surface is low, the TMR decays very fast as the NM thickness
increases. A free-electron case was also considered. The NM band was
assumed to be the same as the majority spin band of the FM layer,
and spin-splitting in the FM layer was assumed to be rather large.
The TMR approached a finite value as the NM thickness increased.
This was because a large portion of electrons with the majority spin
in the left FM layer were reflected completely at the right NM/FM
interface. When the semi-infinite FM layer on the right side was
replaced by a finite layer, the TMR decayed to zero as the NM
thickness increased. As long as the transmission into the FM layer
was higher than that through the tunnel barrier, the TMR became zero
as the thickness of the NM spacer increased. We suggest that finite
TMR for the infinite NM spacer is unrealistic in real experiments.

\begin{acknowledgments}

This work was supported by the Basic Science Research Program
through the National Research Foundation of Korea (NRF), funded by
the Ministry of Education, Science and Technology (Grants No.
2011-0005251 and No. 2012-0002984).

\end{acknowledgments}

\newpage

\begin{figure}[t!]
\includegraphics*[viewport=100 200 600 2000,width=15cm,angle=-90]{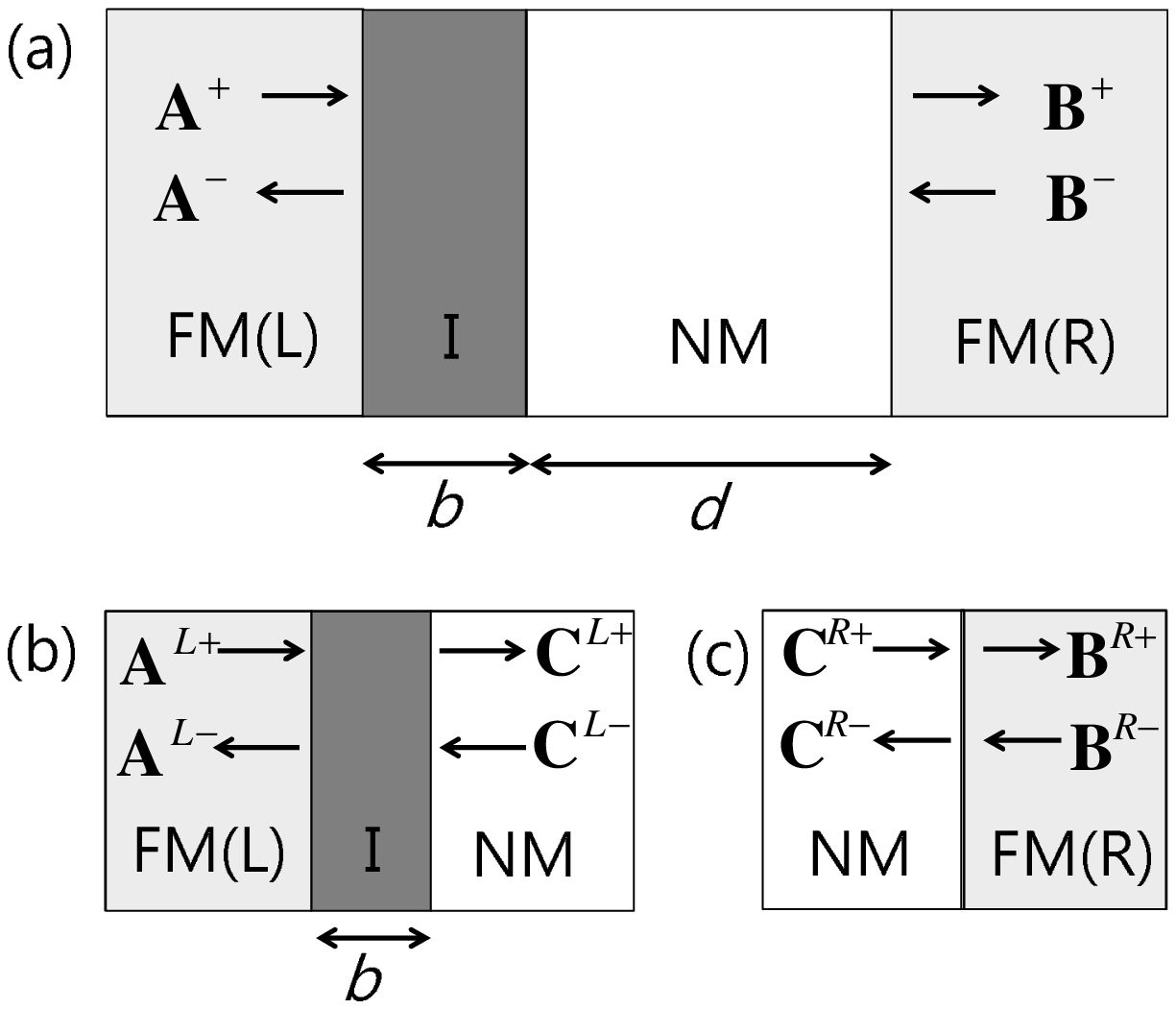}
\caption{(a) Schematic diagram of a magnetic tunnel junction (MTJ)
with a nonmagnetic (NM) layer inserted between the insulating (I)
tunnel barrier and the right ferromagnetic layer [FM(R)]. $d$ is the
thickness of the NM layer. ${\bf A}^+$, ${\bf A}^-$, ${\bf B}^+$,
and ${\bf B}^-$ are coefficient vectors [see Eq.~(\ref{eq:MTJ})].
The transmission in the MTJ can be expressed with the reflection and
transmission amplitudes of the separated interfaces shown in (b) and
(c). }
 \label{fig.1}
\end{figure}

\newpage

\begin{figure}[t!]
\includegraphics*[width=15cm,angle=-90]{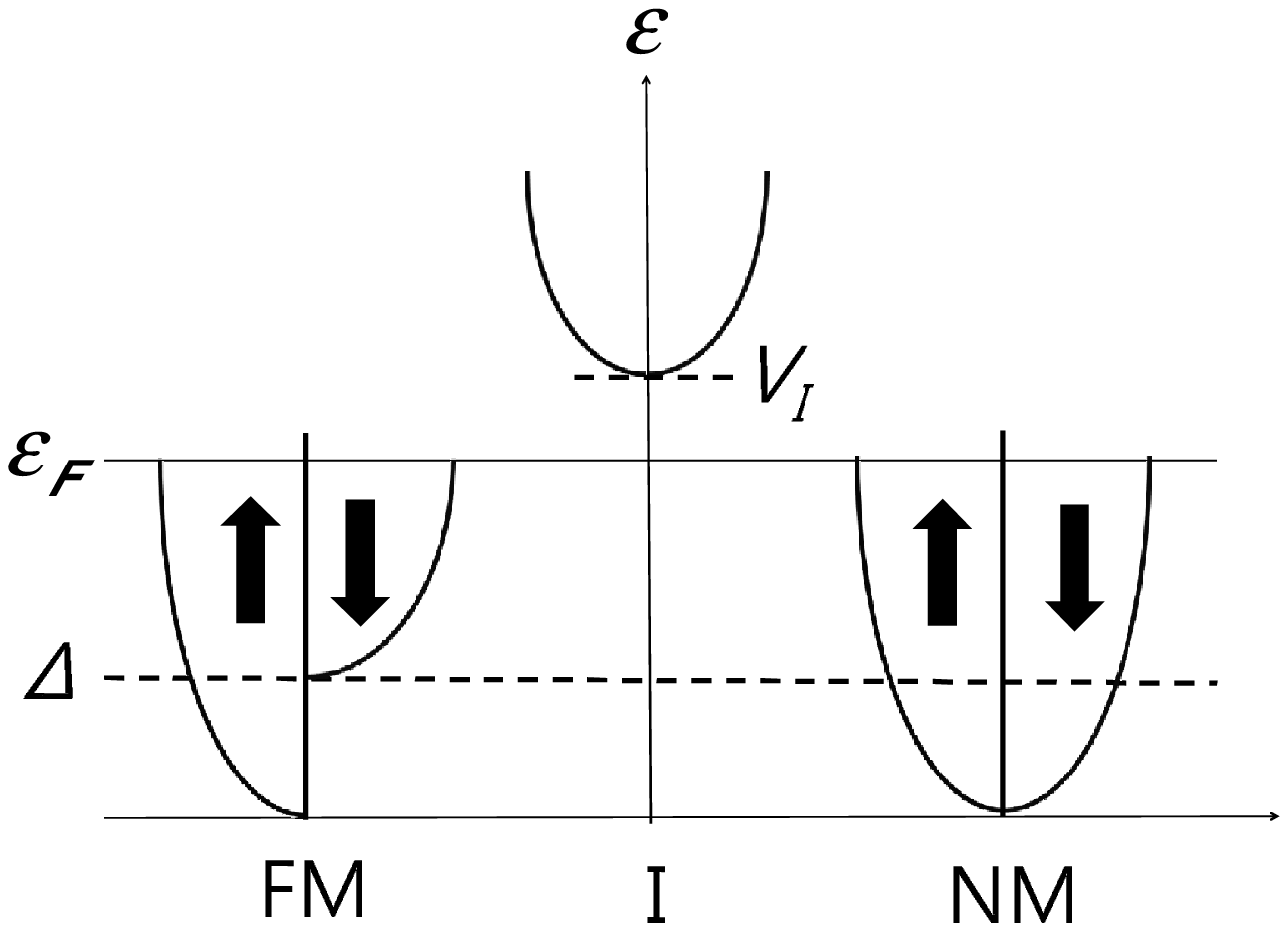}
\caption{Dispersion relation for the FM, I, and NM materials.
$\Delta$ is the spin splitting inside the FM layer, $V_I$ is the
bottom energy for the tunnel barrier, and $\varepsilon_F$ is the
Fermi energy. }
 \label{fig.2}
\end{figure}

\newpage

\begin{figure}[t!]
\includegraphics[width=7cm,angle=-90]{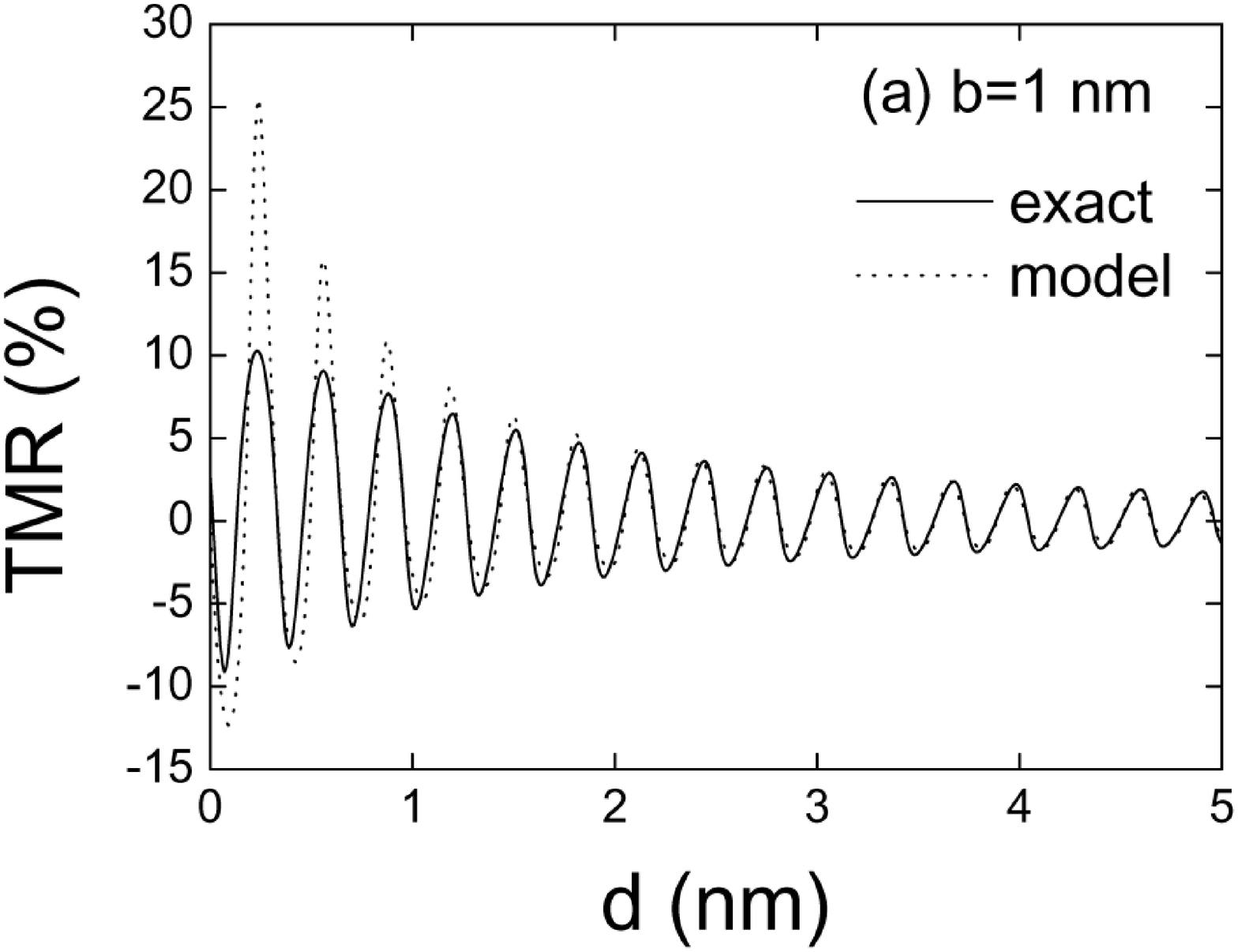}  \\
\includegraphics[width=7cm,angle=-90]{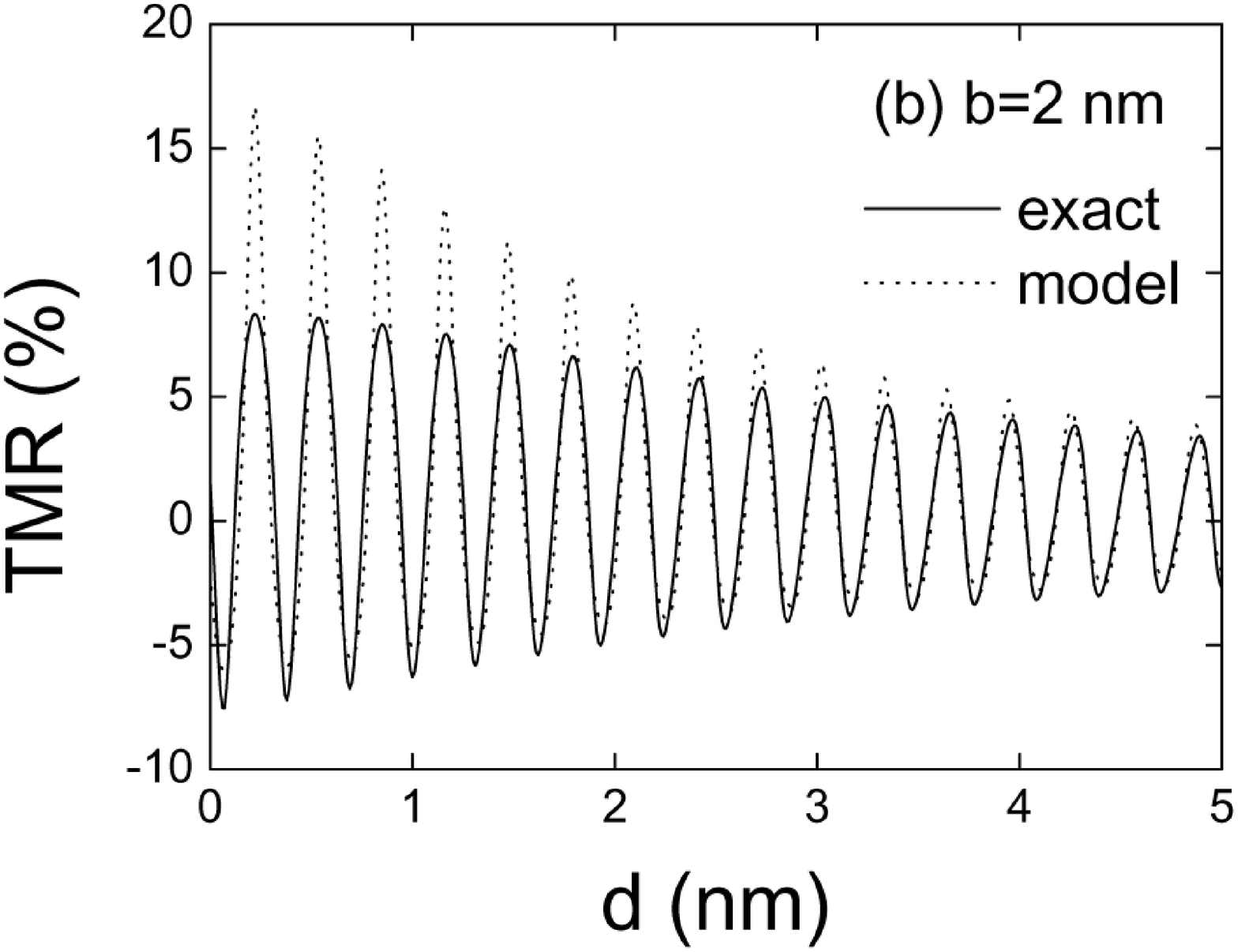}
\caption{TMR as a function of $d$ for the effective-mass band. The
barrier thicknesses are (a) $b=1$ and (b) $b=2$ nm. The parameters
used in the calculation are $\varepsilon_F = 4$ eV, $V_I = 6$ eV,
and $\Delta = 2.5$ eV. The effective mass of electrons with minority
spin in the FM material is $m^*_\downarrow = m_0 \varepsilon_F /
(\varepsilon_F - \Delta )$ and other effective masses are the bare
electron mass $m_0$. The solid line is the exact result, and the
dotted line is based on the proposed analytical formula.   }
\label{EM}
\end{figure}

\newpage

\begin{figure}[t!]
\includegraphics[width=7cm,angle=-90]{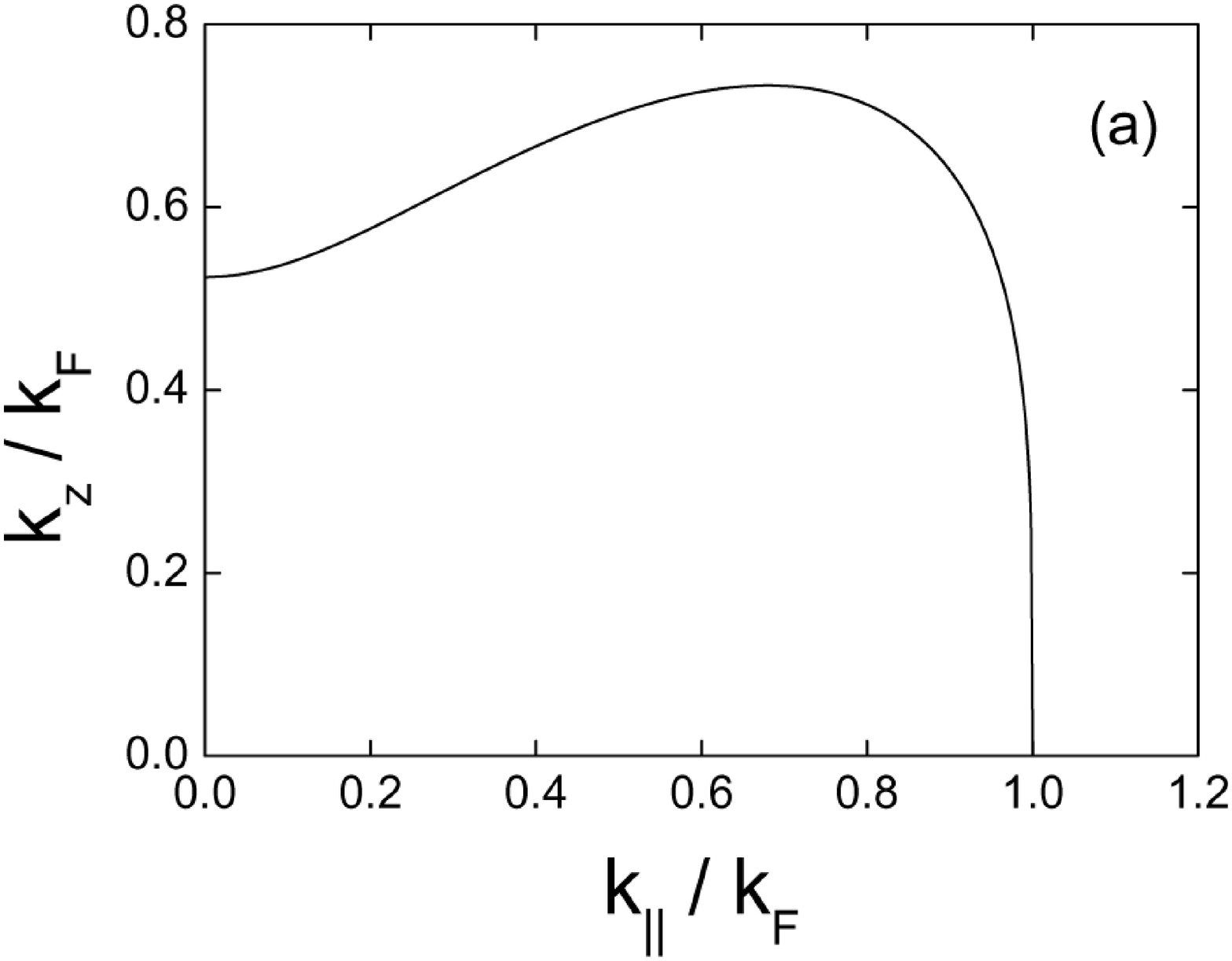}   \\
\includegraphics[width=7cm,angle=-90]{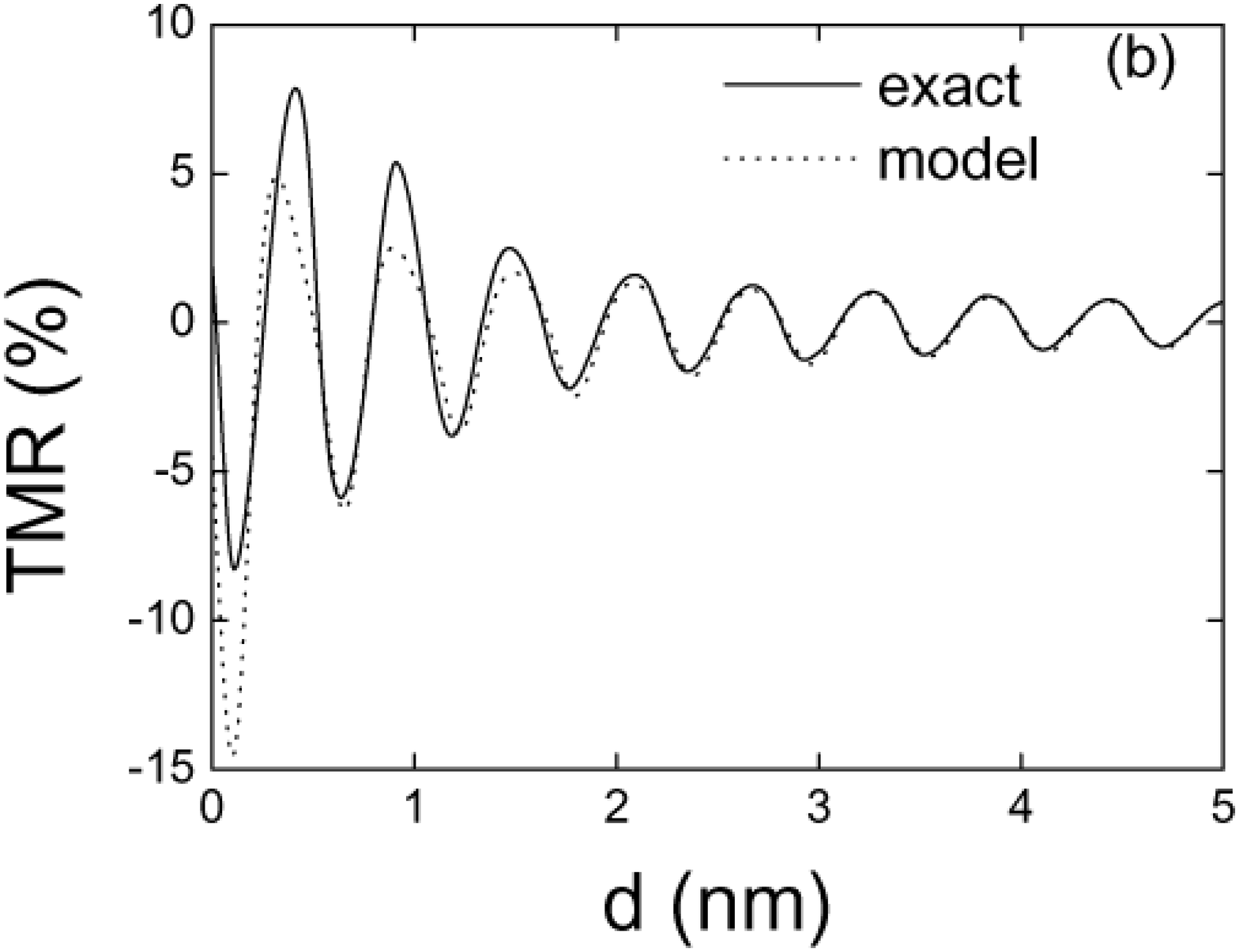}
\caption{(a) Cross section of the Fermi surface for the NM spacer
with the dispersion relation given in Eq.~(\ref{e:NM}). $k_F =
\sqrt{2m_0 \varepsilon_F}/\hbar$ is the Fermi wave vector for the
majority spin of the FM layer. It is similar to the Cu(001) case.
(b) TMR as a function of the NM layer thickness $d$. Except for the
NM spacer, the other parameters are the same as those used in the
previous case. The solid line is the exact result, and the dotted
line is based on the proposed analytical formula. } \label{NM}
\end{figure}

\newpage

\begin{figure}[t!]
\includegraphics[width=7cm,angle=-90]{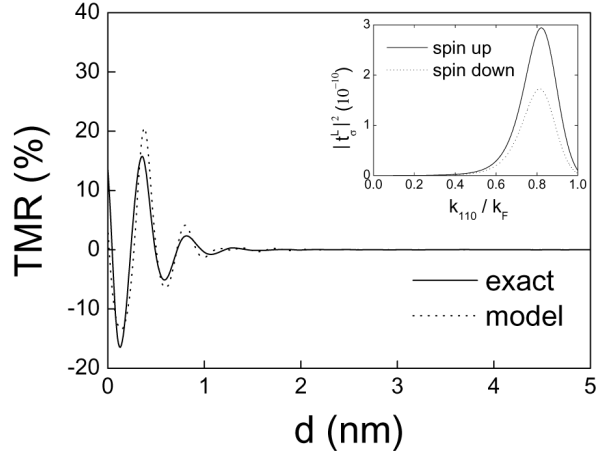}
\caption{TMR as a function of the NM layer thickness $d$ when the
point for the maximum transmission coefficient of the tunnel barrier
does not coincide with the extremal point of the NM Fermi surface in
${\rm \bold k}_{\|}$ space. The dispersion relation of the tunnel
barrier is given in Eq.~(\ref{e:B}), and the FM and NM materials are
assumed to be the same as in Fig.~3. The inset is the transmission
coefficient as a function of ${\rm \bold k}_{\|}$ along the [110]
orientation for the FM/I/NM system. }
\end{figure}

\newpage

\begin{figure}[t!]
\includegraphics[width=7cm,angle=-90]{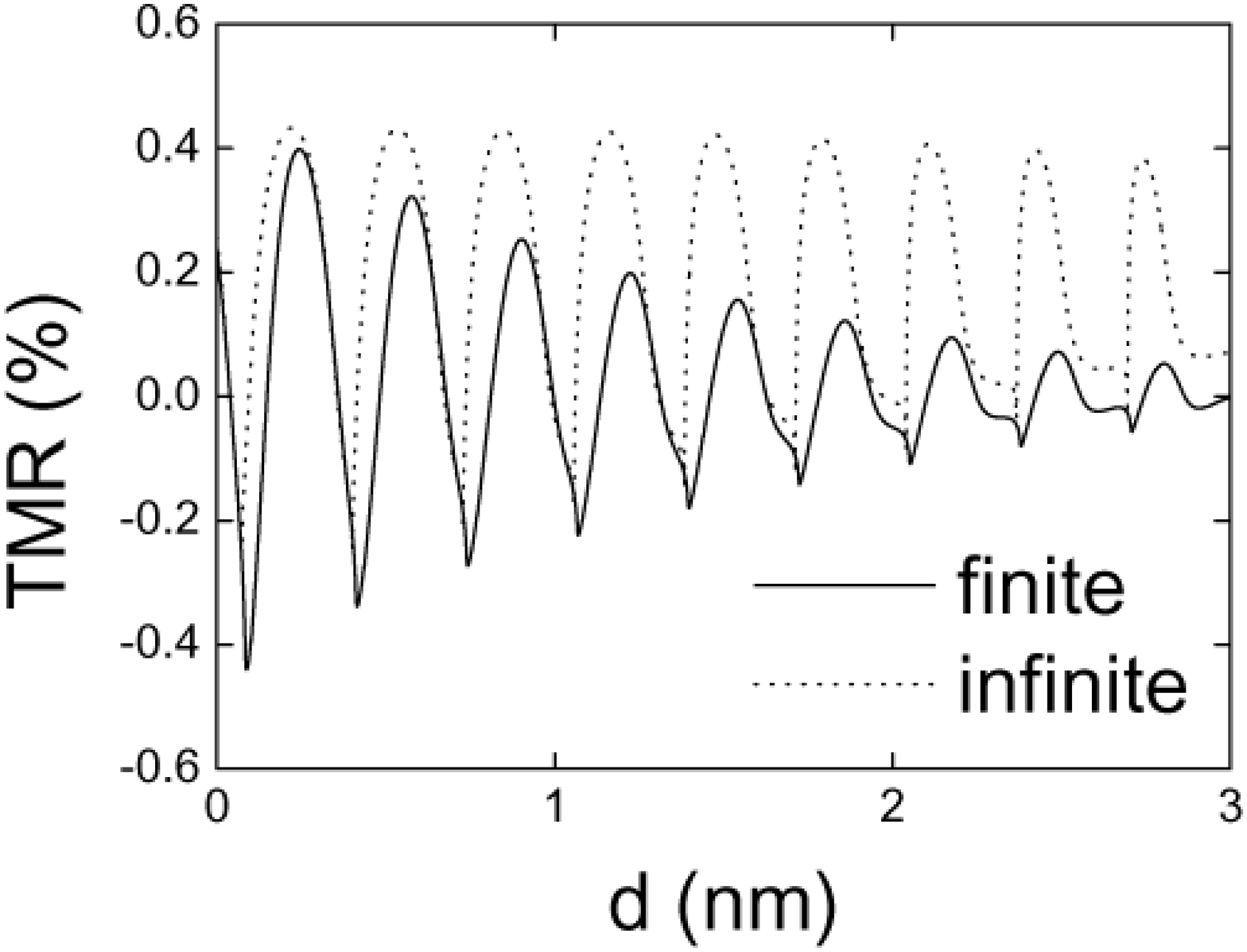}
\caption{TMR as a function of the NM layer thickness $d$ for the
free-electron band. The parameters used in the calculation are
$\varepsilon_F = 4$ eV, $V_I = 6$ eV, $\Delta = 3.5$ eV, and $b=1$
nm. The solid line is for the 1 nm FM layer and the dotted line is
for the infinite FM layer. }
\end{figure}


\begin{thebibliography}{}

\bibitem{mood} J. S. Moodera, L. R. Kinder, T. M. Wong, and R. Meservey,
Phys. Rev. Lett. {\bf 74}, 3273 (1995).

\bibitem{yuasaR} S. Yuasa and D. D. Djayaprawira, J. Phys. D: Appl. Phys.
{\bf 40} R337 (2007).

\bibitem{vedy} A. Vedyayev, N. Ryzhanova, C. Lacroix, L. Giacomoni, and B.
Dieny, Europhys. Lett. {\bf 39}, 219 (1997).

\bibitem{dust} P. LeClair, H. J. M. Swagten, J. T. Kohlhepp,
R. J. M. van de Veerdonk, and W. J. M. de Jonge, Phys. Rev. Lett.
{\bf 84}, 2933 (2000).

\bibitem{dust2} P. LeClair, J. T. Kohlhepp, H. J. M. Swagten, and W. J. M. de
Jonge, Phys. Rev. Lett. {\bf 86}, 1066 (2001).

\bibitem{zhang} S. Zhang and P. M. Levy, Phys. Rev. Lett. {\bf 81},
5660 (1998).

\bibitem{yuasa} S. Yuasa, T. Nagahama, and Y. Suzuki, Science {\bf 297},
234 (2002).

\bibitem{itoh} H. Itoh, J. Inoue, A. Umerski, and J. Mathon, Phys.
Rev. B {\bf 68}, 174421 (2003).

\bibitem{itoh2} H. Itoh, J. Inoue, A. Umerski, and J. Mathon, J.
Magn. Magn. Mater. {\bf 272-276}, e1467 (2004).

\bibitem{shok} A. A. Shokri and A. Saffarzadeh, J. Phys.:Condens. Matter
{\bf 16}, 4455 (2004).

\bibitem{zeng} Z. M. Zeng, X. F. Han, W. S. Zhan, Y. Wang, Z. Zhang, and S. Zhang,
 Phys. Rev. B  {\bf 72}, 054419 (2005).

\bibitem{yang} J. Yang, J. Wang, Z. M. Zheng, D. Y. Xing, and C. R.
Chang, Phys. Rev. B {\bf 71}, 214434 (2005).

\bibitem{niu} Z. P. Niu, Z. B. Feng, J. Yang, and D. Y. Xing,
Phys. Rev. B {\bf 73}, 014432 (2006).

\bibitem{feng} X. Feng, O. Bengone, M. Alouani, I. Rungger, and S. Sanvito,
Phys. Rev. B {\bf 79}, 214432 (2009).

\bibitem{chen} S.-P. Chen and C. R. Chang, IEEE Trans. Magn. {\bf
45}, 2410 (2009).

\bibitem{autes} G. Autes, J. Mathon, and A. Umerski,
Phys. Rev. B {\bf 80}, 024415 (2009).

\bibitem{chen2} S.-P. Chen, J. Appl. Phys. {\bf 107}, 09C716 (2010).

\bibitem{chen3} S.-P. Chen, Thin Solid Films {\bf 519}, 8215 (2011).

\bibitem{lee1} B. C. Lee, J. Appl. Phys. {\bf 107}, 09C708 (2010).

\bibitem{bruu} H. Bruus and K. Flensberg, {\it Many-Body
Quantum Theory in Condensed Matter Physics} (Oxford University
Press, New York, 2004).



\bibitem{lee} B. Lee and Y.-C. Chang, Phys. Rev. B {\bf 52}, 3499
(1995).

\bibitem{lee2} B. C. Lee and Y.-C. Chang, Phys. Rev. B {\bf 62}, 3888  (2000).

\bibitem{stil} M. D. Stiles, J. Magn. Magn. Mater. {\bf 200}, 322
(1999) and references therein.

\bibitem{tse} L. Tsetseris, B. C. Lee, and Y.-C. Chang, Phys. Rev. B
{\bf 55} 11586 (1997).

\bibitem{you} C.-Y. You, C. H. Sowers, A. Inomata, J. S. Jiang, S. D. Bader, and D. D.
Koelling, J. Appl. Phys. {\bf 85}, 5889 (1999).

\bibitem{naga} T. Nagahama, S. Yuasa, E. Tamura, and Y. Suzuki, Phys.
Rev. Lett. {\bf 95}, 086602 (2005).

\bibitem{mats} R. Matsumoto, A. Fukushima, K. Yakushiji, S. Nishioka, T. Nagahama, T. Katayama,
Y. Suzuki, K. Ando, and S. Yuasa, Phys. Rev. B {\bf 79}, 174436
(2009).

%



\end{thebibliography}
\end{document}